\documentclass[12pt,preprint]{aastex}

\newcommand{\civ}{C\,{\sc iv}}
\newcommand{\mgii}{Mg\,{\sc ii}}

\defcitealias{shen08}{S08}
\defcitealias{croom11}{C11}

\begin{document}

\title{The Importance of Broad Emission-Line Widths in Single Epoch
  Black Hole Mass Estimates}

\author{R.J.~Assef\altaffilmark{1},
  S.~Frank\altaffilmark{2,3},
  C.J.~Grier\altaffilmark{2},
  C.S.~Kochanek\altaffilmark{2,4},
  K.D.~Denney\altaffilmark{5},
  B.M.~Peterson\altaffilmark{2,4}
}

\altaffiltext{1} {NASA Postdoctoral Program Fellow at the Jet
  Propulsion Laboratory, California Institute of Technology, MS
  169-530, 4800 Oak Grove Drive, Pasadena, 91109, USA
  [email:{\tt{roberto.j.assef@jpl.nasa.gov}}]}

\altaffiltext{2} {Department of Astronomy, The Ohio State
  University, 140 W.\ 18th Ave., Columbus, OH 43210, USA}

\altaffiltext{3} {Observatoire Astronomique de Marseille-Provence,
  Laboratoire d'Astrophysique de Marseille/LAM, P\^ole de l'\'Etoile Site
  de Ch\^ateau-Gombert 38, rue Fr\'ed\'eric Joliot-Curie 13388 Marseille,
  France}

\altaffiltext{4} {The Center for Cosmology and Astroparticle Physics,
The Ohio State University, 191 West Woodruff Avenue, Columbus, OH
43210, USA}

\altaffiltext{5} {Marie Curie Fellow at the Dark Cosmology Centre,
  Niels Bohr Institute, University of Copenhagen, Juliane Maries Vej
  30, 2100 Copenhagen, Denmark}

\begin{abstract}
Estimates of the mass of super-massive black holes (BHs) in distant
active galactic nuclei (AGNs) can be obtained efficiently only through
single-epoch spectra, using a combination of their broad emission-line
widths and continuum luminosities. Yet the reliability and accuracy of
the method, and the resulting mass estimates, $M_{\rm BH}$, remain
uncertain. A recent study by Croom using a sample of SDSS, 2QZ and
2SLAQ quasars suggests that line widths contribute little information
about the BH mass in these single-epoch estimates and can be replaced
by a constant value without significant loss of accuracy. In this
Letter, we use a sample of nearby reverberation-mapped AGNs to show
that this conclusion is not universally applicable. We use the bulge
luminosity ($L_{\rm Bulge}$) of these local objects to test how well
the known $M_{\rm BH} - L_{\rm Bulge}$ correlation is recovered when
using randomly assigned line widths instead of the measured ones to
estimate $M_{\rm BH}$. We find that line widths provide significant
information about $M_{\rm BH}$, and that for this sample, the line
width information is just as significant as that provided by the
continuum luminosities. We discuss the effects of observational biases
upon the analysis of Croom and suggest that the results can probably
be explained as a bias of flux-limited, shallow quasar samples.
\end{abstract}


\section{Introduction}\label{sec:intro}

Super-massive black holes (BHs) at the center of galaxies are believed
to play a fundamental role in the evolution of galaxies. Accreting
BHs, or active galactic nuclei (AGNs), are thought, for example, to be
responsible for the quenching of star-formation in galaxies needed to
explain the existence of the blue cloud and the red sequence
\citep[e.g.,][]{granato04,dimatteo05,hopkins05} and for heating gas
near the centers of galaxy clusters to stop gas accretion onto the
central galaxies \citep[e.g.,][]{croton06}. This makes characterizing
the masses and accretion rates of the BH population across cosmic time
crucial to understanding the evolution of galaxies and clusters.

Direct dynamical measurements of the mass of BHs are, however,
possible only for nearby quiescent galaxies using spatially resolved
velocity measurements inside (or close to) the black hole's sphere of
influence \citep[e.g.,][]{merritt01,gultekin09}. Fortunately, AGNs
provide a completely different means of directly estimating $M_{\rm
  BH}$ through reverberation mapping
\citep[RM;][]{blandford82,peterson93}. Here it is assumed that the gas
responsible for the broad-line emission is virialized and that the
broad-line velocity widths are related to the orbital speed of the gas
around the black hole. Estimates of $M_{\rm BH}$ are then obtained by
combining the width of the emission lines with an estimate of the
distance from the BH to the broad line region ($R_{\rm BLR}$). Since
AGNs are typically variable, $R_{\rm BLR}$ is estimated using the time
it takes for the broad-emission lines to respond to changes in the
accretion disk luminosity.

While the RM technique can in principle be used for sources at any
distance, it is time consuming, and has been impractical for luminous
and distant objects whose variability timescales are long due to their
high luminosity \citep[e.g.,][]{vandenberk04,macleod10} and time
dilation. Local RM observations have, however, shown that $R_{\rm
  BLR}$ is tightly correlated with the luminosity of the accretion
disk \citep[e.g.,][]{kaspi00,bentz09}. This allows one to estimate
$R_{\rm BLR}$ directly from the luminosity, and hence estimate $M_{\rm
  BH}$ from the single-epoch (SE) spectra used to determine the line
widths. In practice, SE $M_{\rm BH}$ estimates are obtained by means
of
\begin{equation}\label{eq:mbh_se}
M_{\rm BH}\ =\ f\ \frac{(\Delta v)^2\ L_{\lambda}^{\alpha}}{G}
\end{equation}
\noindent where $\Delta v$ is the velocity width of a given broad
emission line, $L_{\lambda}$ is the luminosity of the continuum at an
associated wavelength $\lambda$ (in \AA), $\alpha\approx 1/2$ but may
depend weakly on $\lambda$, $G$ is the gravitational constant and $f$
is a dimensionless factor that accounts for the geometry and
inclination of the BLR and the characterization of the line width used
in defining the relation. The two most common line-width
characterizations are the full-width at half maximum (FWHM) and the
line dispersion \citep[$\sigma_l$; see, e.g.,][]{peterson04}. The most
common combinations of broad lines and continuum luminosities used are
H$\beta$ and $L_{5100}$ \citep[$\alpha = 0.519$,][]{bentz09} at low
redshift, and \mgii\ with $L_{3000}$ \citep[$\alpha =
  0.47$,][]{mclure02} or \civ\ with $L_{1350}$ \citep[$\alpha =
  0.53$,][]{vestergaard06} at higher redshifts, where these lines
appear at observed-frame optical wavelengths.

Recently, \citet[][hereafter \citetalias{croom11}]{croom11} used a
large sample of quasars from the Sloan Digital Sky Survey
\citep[SDSS,][]{york00}, the 2dF QSO Redshift Survey
\citep[2QZ,][]{croom04} and the 2dF-SDSS LRG and QSO Survey
\citep[2SLAQ,][]{richards05} to study the importance of line-width
estimates for the accuracy of SE $M_{\rm BH}$ estimates. For each of
the H$\beta$, \mgii, and \civ\ lines, \citetalias{croom11} observed
that the distributions of SE $M_{\rm BH}$ estimates were not
significantly different before and after scrambling the line widths
across the sample. \citetalias{croom11} also noted that the
distribution of line widths with redshift was relatively narrow and
showed little evolution with redshift. \citetalias{croom11} concluded
that line widths provide little additional information about BH masses
as required by equation (\ref{eq:mbh_se}) compared to simply using a
constant value.

While this is a very interesting observation, it conflates three
possible explanations: i) that the underlying assumptions behind
equation (\ref{eq:mbh_se}) are not correct and that $\Delta v$ truly
holds no physical information, ii) that for the typically low $S/N$
spectra used, estimates of $\Delta v$ are so noisy that their physical
information is lost or biased, and iii) that due to a conspiracy
between the survey selection function and the evolution of the quasar
mass, luminosity and accretion rate distributions, the observed
distribution of $\Delta v$ is narrow and non-evolving with redshift,
minimizing the effects of its randomization on $M_{\rm BH}$. In this
Letter, we try to separate these issues. First, in \S\ref{sec:test},
we consider the local, reverberation-mapped sample, where independent
indirect estimates of $M_{\rm BH}$ are possible through the well-known
correlation with the luminosity of the spheroidal component of the
host galaxy. Using this sample, we show that the velocities provide
almost as much information on $M_{\rm BH}$ as the luminosities. Then,
in \S\ref{sec:discussion} we discuss the effects of observational
biases on the conclusions of \citetalias{croom11}.  Where needed, we
assume a $\Lambda$CDM flat cosmology with $H_0 = 73~\rm km~\rm
s^{-1}~\rm Mpc^{-1}$, $\Omega_{\Lambda}=0.7$ and $\Omega_M = 0.3$.

\section{Are Line Widths Important for Accurate BH Mass Estimates?}\label{sec:test}

In order to understand the importance of line widths in SE $M_{\rm
  BH}$ estimates, we use a sample of 34 nearby AGNs observed in
reverberation mapping (RM) campaigns with bulge luminosity
measurements at 5100\AA, $L_{\rm Bulge}$, by \citet{bentz09c} using
{\it{HST}}. Although the target selection is somewhat arbitrary, the
sample is approximately volume limited, as luminous, distant sources
are generally avoided in RM studies due to cosmological time-dilation
and their intrinsically longer variability timescales. The inherent
difficulty in measuring bulge luminosities of distant AGNs further
constrains the sample distance. The sample properties are summarized
in Table \ref{tab:sample}.

Every object in our sample has been the subject of at least one RM
observational campaign, and many for two or more. For each object we
use the mean optical spectrum of each reverberation mapping campaign,
and we treat each of the 62 mean spectra as independent objects. While
there is typically not enough variability from a single object within
an RM campaign to treat each individual spectrum as an independent
data point, it is a good approximation to assume each observational
campaign provides an independent measurement. Although for any one
object the dynamic range in luminosity is limited, \citet{peterson04}
and \citet{bentz10} show that RM $M_{\rm BH}$ estimates of NGC~5548,
the most studied source, are consistent between campaigns ($\Delta
\log{M_{\rm BH}}\approx 0.14~\rm dex$). Table \ref{tab:sample} also
lists the number of individual spectra used to build each mean
spectrum.

\begin{figure}
  \begin{center}
    \plotone{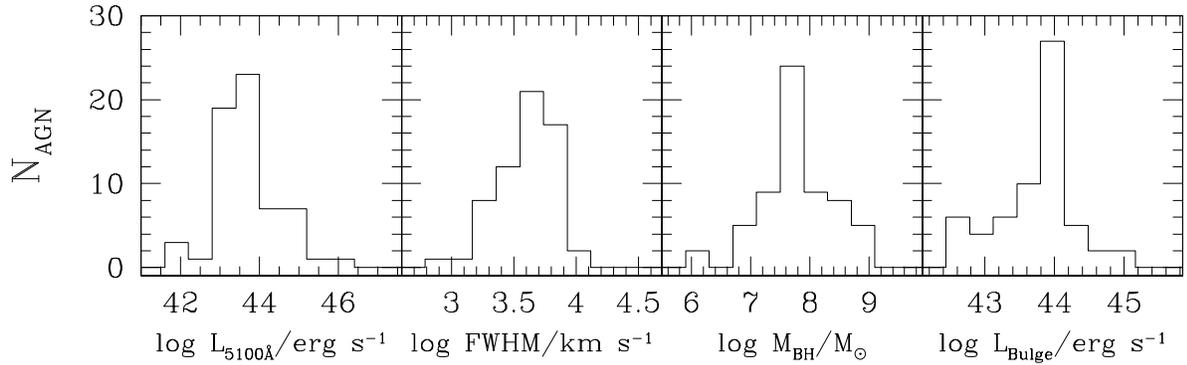}
    \caption{Distribution of the physical properties of our sample of
      nearby AGNs. The panels show the distribution of the $L_{5100}$
      AGN continuum luminosities ({\it{left}}), broad H$\beta$ FWHM
      ({\it{middle left}}), SE $M_{\rm BH}$ estimates ({\it{middle
          right}}) and 5100\AA\ bulge luminosities $L_{\rm Bulge}$
      ({\it{right}}).}
    \label{fg:sample}
  \end{center}
\end{figure}

The host-corrected AGN continuum luminosities at 5100\AA, $L_{5100}$,
are taken from \citet{bentz09c}, while the broad H$\beta$ FWHM and
$\sigma_l$ measurements are taken directly from the references listed
in Table \ref{tab:sample}. While this leads to some heterogeneity in
the recipes used to make the line-width measurements, the (typically)
extremely high $S/N$ of the mean spectra minimizes the resulting
systematic errors \citep{denney09}. The distributions of the sample in
H$\beta$ FWHM, continuum $L_{5100}$, bulge $L_{\rm Bulge}$, and
estimated black hole mass are shown in Figure \ref{fg:sample}.

The ideal test to determine whether the velocity widths hold
information about the BH mass would compare these single-epoch mass
estimates to dynamical $M_{\rm BH}$ measurements for the same
objects. Unfortunately, few AGNs exist with dynamical measurements of
$M_{\rm BH}$, rendering such a test infeasible at present. Studies of
quiescent galaxies, however, have determined that there is a strong
correlation between the mass of the central BH and the luminosity of
the spheroidal component of its host galaxy
\citep[e.g.,][]{marconi03,graham07,graham12}. So while we lack
dynamical estimates of $M_{\rm BH}$ for comparison, we can assess the
reliability of SE $M_{\rm BH}$ estimates simply by using $L_{\rm
  Bulge}$ as a proxy and studying how well they hold to the $M_{\rm
  BH} - L_{\rm Bulge}$ correlation.

We estimate SE BH masses for each object by means of equation
(\ref{eq:mbh_se}), using the mean continuum luminosity at 5100\AA\ and
either the FWHM or the $\sigma_l$ line width characterization of the
mean H$\beta$ spectrum. We evaluate the correlation strength between
$M_{\rm BH}$ and $L_{\rm Bulge}$ using the Spearman rank-order
coefficient, $r_s$, finding a value of $r_s^0 = 0.69~(0.63)$ when
using the FWHM ($\sigma_l$) for the mass estimates. Given the number
of data points, the probability of obtaining such a value of $r_s^0$
in the absence of a correlation is $P_{\rm Ran} = 7.4\times
10^{-10}~(5.4\times 10^{-8})$. As expected, our SE $M_{\rm BH}$
estimates are very well correlated with the luminosity of the
spheroidal component of their host, suggesting these SE estimates can
be quite accurate.

\begin{figure}
  \begin{center}
    \plotone{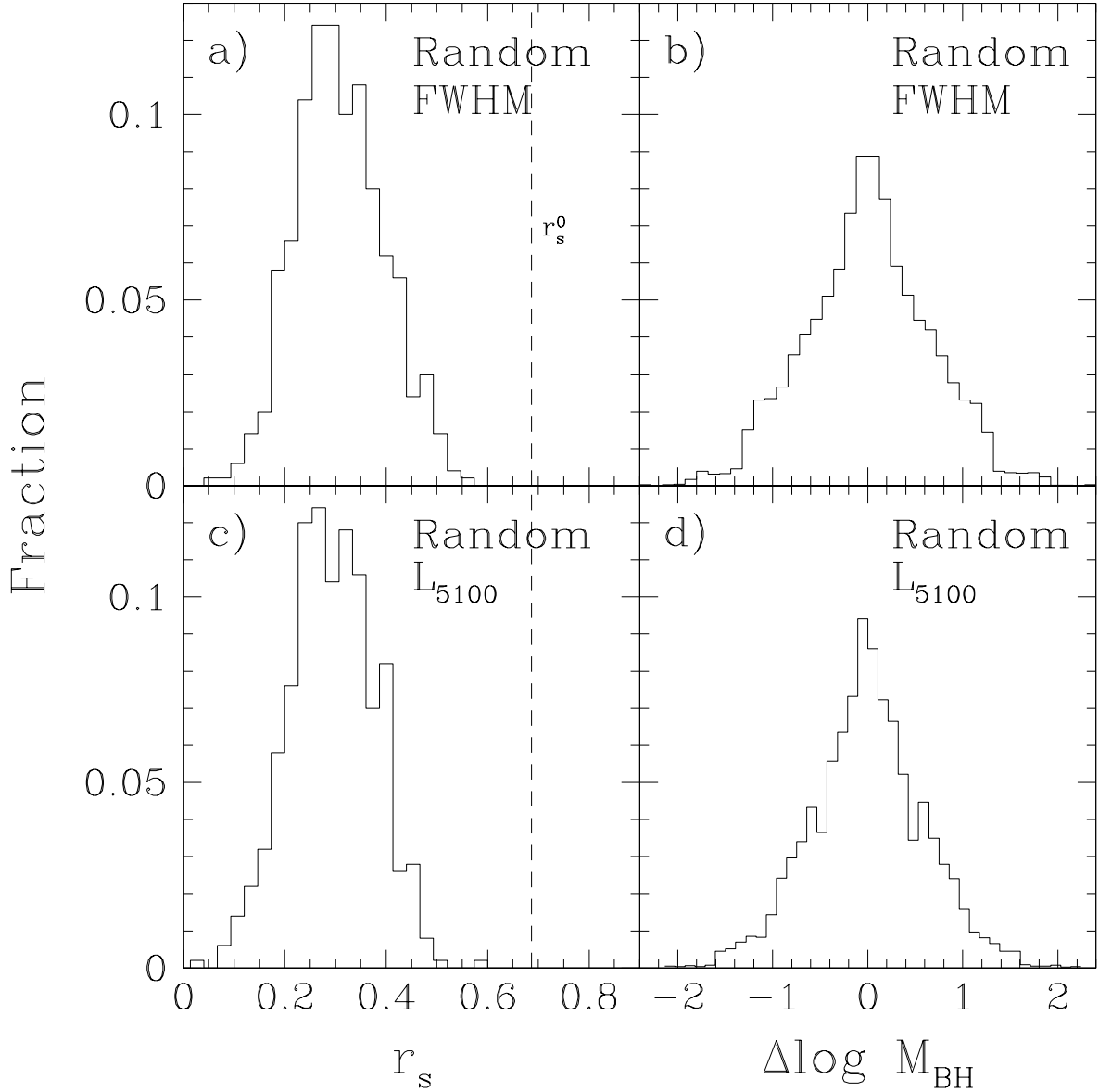}
    \caption{Panel {\it{a}}) shows the distribution of Spearman
      rank-order coefficients $r_s$ obtained for 500 randomizations of
      the FWHM measurements across our sample. The dashed line shows
      the value obtained using the FWHM measurement that corresponds
      to each object. Panel {\it{b}} shows the distribution of the
      changes in $M_{\rm BH}$ caused by randomizing the line-widths,
      i.e., $\Delta \log M_{\rm BH} = \log M_{\rm BH} ({\rm FWHM_{\rm
          Random}}) - \log M_{\rm BH} (\rm FWHM_{\rm True})$. The
      histogram is constructed by combining the logarithmic change in
      $M_{\rm BH}$ for every object for every realization. Panels
      {\it{c}}) and {\it{d}}) are analogs to panels {\it{a}}) and
      {\it{b}}) but after randomizing $L_{5100}$ instead of FWHM.}
    \label{fg:spear_rand}
  \end{center}
\end{figure}

The amount of information obtained by using accurate line-width
estimates can then be tested by re-estimating $r_s$ using $M_{\rm BH}$
estimates obtained with randomly selected values for the line
widths. In practice, we randomly re-distribute the measured line
widths across the objects in our sample 500 times, essentially
bootstrap resampling the observed FWHM distribution, estimating $r_s$
for each realization. In this procedure we never assign a measurement
its true line width, although doing so does not alter our main
conclusions. Figure \ref{fg:spear_rand} shows the resulting
distribution of the $r_s$ coefficients compared to the value obtained
before modifying the original line widths, $r_s^0$. We only show the
results for the FWHM, as those using $\sigma_l$ are very similar. Note
that $r_s^0$ is significantly above the distribution obtained by
randomizing the line widths. If we view the values of $r_s$ as
Gaussian random numbers, the true value is 4.3$\sigma$ from the mean
of the random trials. The mean value of $r_s = 0.309$ ($P_{\rm Ran} =
1.5\%$) in the random trials is 8 orders of magnitude less significant
than that obtained using the real FWHM estimates. This demonstrates
that line widths provide very significant information about $M_{\rm
  BH}$ and that accurate line-width estimates are crucial for accurate
SE $M_{\rm BH}$ estimates in this sample. This is illustrated in the
top-right panel of Figure \ref{fg:spear_rand}, which shows the spread
in BH mass ratio obtained by combining all 500 random
realizations. The difference in the estimated $M_{\rm BH}$ is strongly
peaked but also with broad wings, with a standard deviation of
0.65~dex. We find similar results if we include each source only once.

Using the same formalism, we can also investigate the importance of
the continuum luminosity for SE $M_{\rm BH}$ estimates by randomizing
the continuum luminosities rather than the line-width estimates. The
resulting $r_s$ distribution is also shown in Figure
\ref{fg:spear_rand}, and it looks very similar to that obtained by
randomizing the line widths. This is not surprising, as in our sample
the $\sim$2~dex dynamic range in $L_{\rm 5100}^{1/2}$ is similar to
that of $\rm FWHM^2$ (see Fig. \ref{fg:sample}). The mean Spearman
Rank-order coefficient of the randomized distributions is $r_s=0.293$
($P_{\rm Ran}=2.1\%$) so, for this sample, accurate line widths are
just as important as accurate luminosities for SE $M_{\rm BH}$
estimates. Finally, as a cross-check, one can also estimate the
probability that the $M_{\rm BH} - L_{\rm Bulge}$ correlation could
randomly arise in our sample. We randomize the sample in both
$L_{5100}$ and the FWHM, and, as expected, we find an $r_s$
distribution consistent with no correlation. The mean of the
distribution is $r_s = -1.7\times 10^{-3}$, implying $P_{\rm Ran} =
99\%$.

\section{Discussion}\label{sec:discussion}

Unless the results of the previous section are not applicable at
higher redshift, which is physically implausible, we must now look
into the possibility that \citetalias{croom11}'s results are either
due to the low $S/N$ of the spectra or a conspiracy between the
surveys and the shape and evolution of both the QSO luminosity and
Eddington ratio distributions that leaves the observed line widths in
a nearly redshift independent distribution. We note that the sample of
\citetalias{croom11} is overwhelmingly dominated by SDSS observations,
accounting for 100\%, 71\% and 86\% of the H$\beta$, \mgii\ and
\civ\ measurements, respectively. For simplicity, the following
discussion focuses solely on the SDSS data.

\citet{denney09} have shown that H$\beta$ line-width estimates can be
significantly in error for low signal-to-noise ratio ($S/N$) spectra,
with systematic and random error components that scale with $S/N$. The
errors are small for $S/N \sim 20$, but at a typical $S/N \sim 5$,
\citet{denney09} observed a shift of 0.10~dex to larger BH masses, and
the dispersion increased by 0.12~dex. While in principle this shift
could be due to either errors in the line widths or continuum
luminosities, $L_{5100}$ is quite insensitive to $S/N$ and $M_{\rm
  BH}$ only depends on its square root (eqn. [\ref{eq:mbh_se}]). Thus,
the errors in $M_{\rm BH}$ found by \citet{denney09} should be
dominated by uncertainties in the line-width measurements induced by
poor $S/N$.

Given these results, we then consider whether the SDSS data are simply
too noisy. We investigated this possibility using a stacking analysis
of the SDSS QSO spectra. We use \civ\ because it has a clean local
continuum, and is the line most affected by systematic issues in low
$S/N$ spectra \citep[see, e.g.,][]{assef11}. While this is not ideal,
since our discussion and the results of \citet{denney09} have focused
solely on H$\beta$, it is still useful since \citetalias{croom11}
reached identical conclusions for all three of the main QSO broad
emission lines used for $M_{\rm BH}$ estimates (H$\beta$, \mgii\ and
\civ). The details and analysis of these stacked spectra, as well as
for \mgii, are presented by \citet{frank11}. The stacks are produced
by averaging approximately 7200 QSO SDSS spectra from the sample of
\citet[][hereafter \citetalias{shen08}]{shen08} in the redshift range
$1.857 < z < 3.118$ and in the small absolute magnitude range $-27.6 <
M_{i} < -26.6$~mag, divided into 33 bins of FWHM as measured by
\citetalias{shen08}. SDSS quasar redshifts are determined from all
possible absorption and emission lines, minimizing the effects of
possible \civ\ blueshifts upon $z$ \citep[see][for
  details]{adelmanmccarthy08}. In the selection process, known BALQSOs
were rejected as they can significantly bias the resulting combined
spectra. Before stacking, each individual spectrum is corrected for
Galactic reddening using the extinction map of \citet{schlegel98},
shifted into the rest frame, and normalized by the flux near a
rest-frame wavelength of $1700\rm\AA$. The resulting spectra are then
continuum subtracted and averaged, and the FWHM is measured directly
from the stacked spectra. Our results are qualitatively insensitive to
whether we use the median or averaged stacked spectra.

\begin{figure}
  \begin{center}
    \plotone{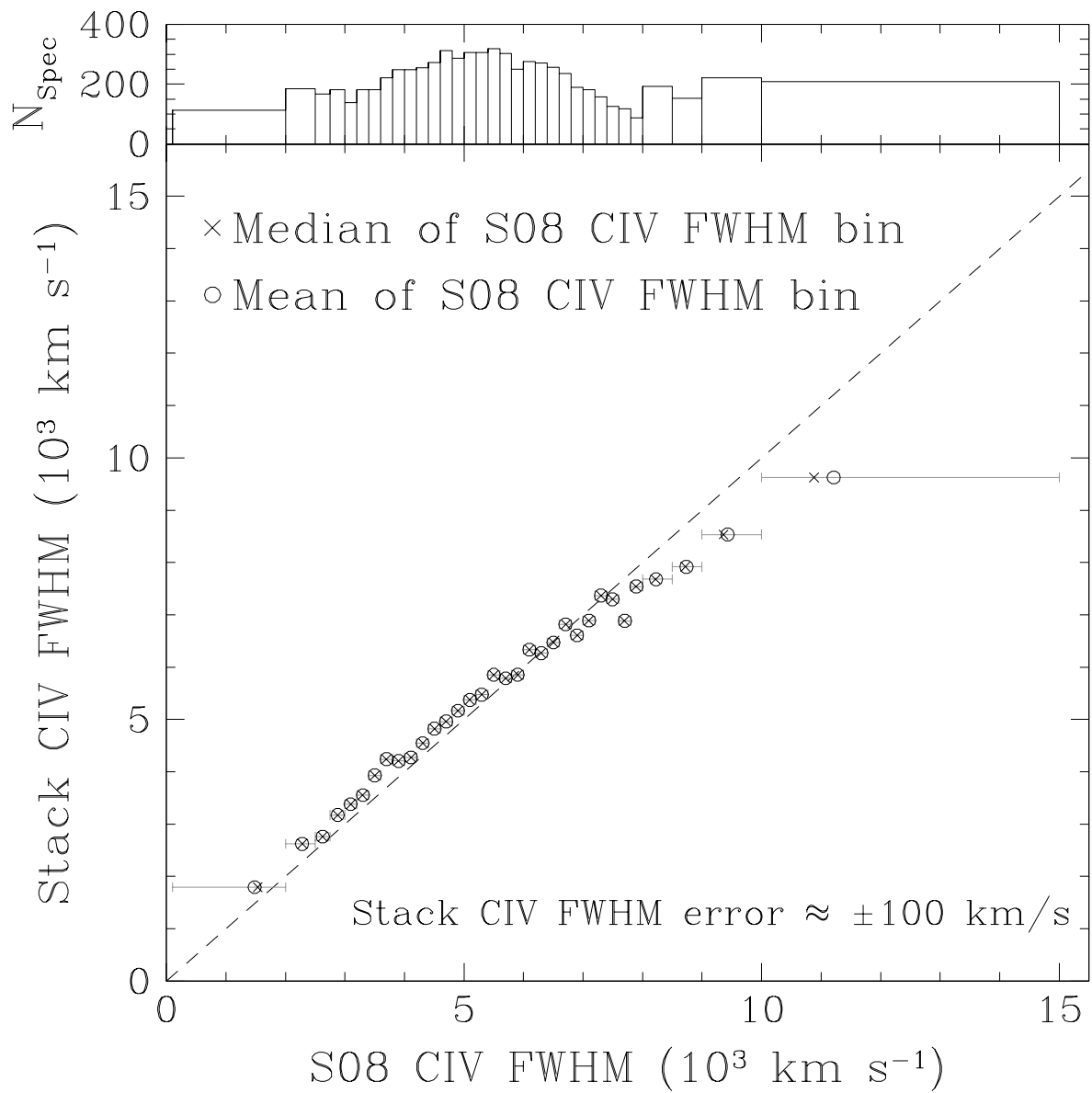}
    \caption{The FWHM of the stacked spectra for each bin of
      individual \citetalias{shen08} FWHM estimates. The open circles
      (crosses) show the mean (median) of the \citetalias{shen08}
      estimates in each bin. The two estimates agree if on the dashed
      line. The mean and median are generally almost equal. The
      horizontal error bars span the range of each \citetalias{shen08}
      FWHM bin. The upper panel shows the number of spectra in each
      \citetalias{shen08} FWHM bin.}
    \label{fg:civ_fwhm_comp}
  \end{center}
\end{figure}

Figure \ref{fg:civ_fwhm_comp} compares the FWHM measured from the
stacked spectra with the mean and median FWHM estimates of
\citetalias{shen08} in each bin. We find that the FWHM of the mean
spectra agree with the mean/median of the individual estimates. Note,
however, that they are moderately biased, particularly at the high
velocity end. These issues are explored further by \citet{frank11}. We
have also repeated the process for different bins of $M_i$, and find
qualitatively similar results for all absolute magnitudes. Overall, we
consider it is unlikely that the line-width measurements are noisy
enough to explain the \citetalias{croom11} results, since on average
they retain the information of the high $S/N$ stacks, and \civ\ is the
line that is most likely to be affected by systematic
biases. Furthermore, we note that \citetalias{croom11} uses the
inter-percentile value to characterize the line widths for \mgii\ and
\civ, and these should be more robust at lower $S/N$ than the FWHM.

\begin{figure}
  \begin{center}
    \plotone{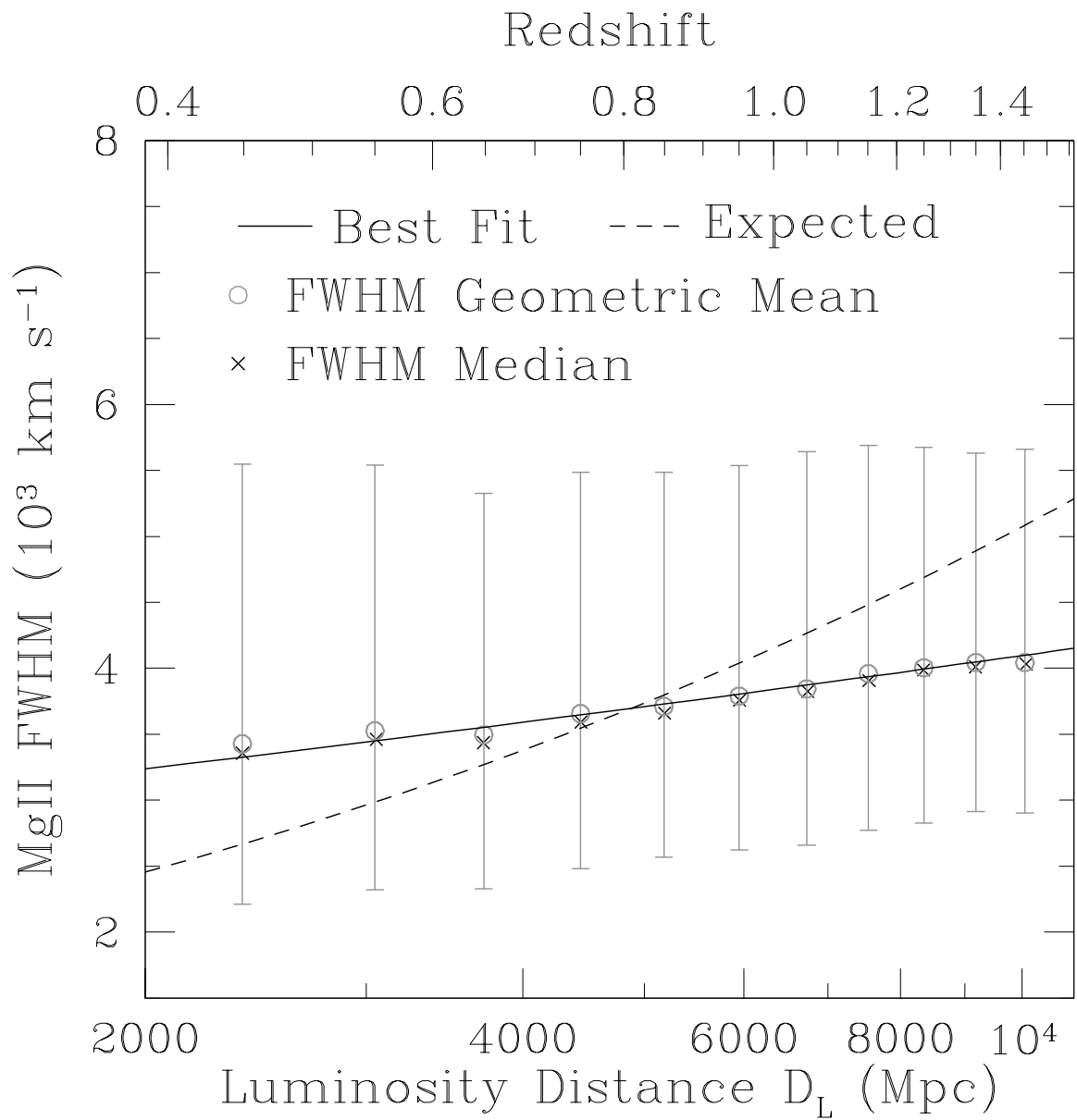}
    \caption{The points show the \citetalias{shen08} Mg\,{\sc ii} FWHM
      as a function of redshift. Gray open circles (crosses) show the
      FWHM geometric mean (median) of the sample in redshift slices of
      width $\Delta z = 0.1$ and the error bars enclose 68.3\% of
      objects above and below the mean. The solid black line shows the
      best-fit evolution of $\rm FWHM \propto \Gamma(z)^{0.16}$ and
      the dashed line shows the expected evolution of
      $\Gamma(z)^{0.5}$ in the absence of $\ell_{Edd}$ evolution,
      normalized to the observed trend at $D_L=5000~\rm Mpc$.}
    \label{fg:fwhm_evo}
  \end{center}
\end{figure}

Thus, we are left with the conclusion that the narrow and apparently
unevolving distribution of line-width measurements in the
\citetalias{croom11} sample (see his Fig. 1) are caused by a
combination of evolution in the Eddington ratio distribution, the
quasar luminosity function (QLF) and the SDSS selection function.  For
example, Figure \ref{fg:fwhm_evo} shows the distribution of
\mgii\ line widths from \citetalias{shen08}, selected because they
span the broadest range of redshifts. With the simplifying
approximation that $\alpha=1/2$ in equation (\ref{eq:mbh_se}), the
line width is given by
\begin{equation}
  \Delta v\ \propto\ {L^{1/4} \over \ell_{Edd}^{1/2}},
\end{equation}
where $\ell_{Edd}$ is the Eddington ratio. We know from
\citet{kollmeier06} that the $\ell_{Edd}$ distribution of broad-line
quasars is narrow. For these bright SDSS AGNs, the steep QLF
\citep[$dn/dL \propto L^{-3.1}$ for $z<2.4$,][]{richards06} means that
at any given redshift half of quasars are within 0.35~mag of the
survey flux limit, corresponding to a velocity spread of only
0.035~dex (8\%) if we neglect the narrow width of the $\ell_{Edd}$
distribution. Thus, at fixed redshift we should expect a narrow
velocity distribution.

The distribution of $\Delta v$ also shows little evolution with
redshift. Note that the luminosity of objects at the flux limit
evolves as $L_{min}\propto \Gamma(z)^2$, where $\Gamma(z) =
D_L(z)\ (1+z)^{-0.28}\sim z^{1.07}$ for our cosmology and a mean
quasar spectral slope of $f_{\nu}\propto \nu^{-0.44}$
\citep{vandenberk01}. Hence, at fixed $\ell_{Edd}$, the apparent
evolution of their associated line widths is $\Delta v \propto
\Gamma(z)^{1/2}\sim z^{0.54}$. If we compare this ``passive''
evolution rate to that observed for \mgii\ (Fig.~\ref{fg:fwhm_evo}),
we find that the differences are already surprisingly small. Taken at
face value, the difference between the ``passive'' evolution model and
the observed $\rm FWHM\propto \Gamma^{0.16}$ implies an evolution in
the Eddington ratio of $\ell_{Edd}\propto \Gamma^{0.68}\sim
z^{0.73}$. Hence, the slow evolution of $\Delta v$ may be related to
quasar ``downsizing'' \citep[e.g.,][]{cowie03}. Whether the slow
evolution is due to general changes in the accretion rates with $z$ or
evolution in $\ell_{Edd}$ as a function of $M_{\rm BH}$ is a complex
problem beyond the scope of this paper. Recent modeling of the QLF and
observed $\ell_{Edd}$ distributions by \citet{shankar11} suggests that
$\ell_{Edd}$ likely evolves in both manners, increasing with
increasing $z$ and decreasing $M_{\rm BH}$. That the net result would
be to leave almost no apparent evolution is then simply a
coincidence. Adding the significant noise in the line-width
measurements of low $S/N$ spectra may further blur the signs of a weak
evolution with redshift. Since line widths are clearly essential for
SE $M_{\rm BH}$ estimates in the local reverberation mapped sample,
this seems to be the most natural explanation of the
\citetalias{croom11} result. We conclude that line widths are as
important as continuum luminosities for determining accurate BH
masses, but that the evolution of the QSO population likely conspires
with the survey flux limit to conceal this in shallow surveys such as
SDSS.

\acknowledgements{We thank David H. Weinberg, Daniel K. Stern, Alister
  Graham and Scott Croom for comments and suggestions that helped
  improve our work. CSK is supported by NSF grant AST-1004756. BMP and
  CJG are grateful from support by NSF grant AST-1008882.}

\begin{deluxetable}{l c c c c c c}

\tablecaption{Reverberation Mapping Sample\label{tab:sample}}

\tablehead{
  \multicolumn{1}{l}{Object}& 
  \colhead{$\log L_{5100\AA}$}&
  \colhead{$\log L_{\rm Bulge}$}&
  \colhead{FWHM}&
  \colhead{$\sigma_l$}&
  \colhead{$\rm N_{\rm Spec}$}&
  \colhead{Ref.}\\
  \colhead{}&
  \colhead{$\rm erg~\rm s^{-1}$}&
  \colhead{$\rm erg~\rm s^{-1}$}&
  \colhead{$\rm km~\rm s^{-1}$}&
  \colhead{$\rm km~\rm s^{-1}$}&
  \colhead{}&
  \colhead{}
}
\tabletypesize{\small}
\tablewidth{0pt}
\tablecolumns{7}

\startdata
Mrk335                              &   43.73 $\pm$ 0.05 &   43.14 & \phn  1792   $\pm$\phn\phn\phn   3 &\phn  1380   $\pm$\phn\phn\phn   6 &  123 &  1\\ 
                                    &   43.81 $\pm$ 0.05 &   43.14 & \phn  1679   $\pm$\phn\phn\phn   2 &\phn  1371   $\pm$\phn\phn\phn   8 &   25 &  1\\ 
PG0026+129                          &   44.95 $\pm$ 0.08 &   44.36 & \phn  2544   $\pm$\phn\phn  56 &\phn  1738   $\pm$\phn 100 &   53 &  1\\ 
PG0052+251                          &   44.78 $\pm$ 0.09 &   44.12 & \phn  5008   $\pm$\phn\phn  73 &\phn  2167   $\pm$\phn\phn  30 &   56 &  1\\ 
F9                                  &   43.94 $\pm$ 0.06 &   44.10 & \phn  5999   $\pm$\phn\phn  60 &\phn  2347   $\pm$\phn\phn  16 &   29 &  1\\ 
Mrk590                              &   43.55 $\pm$ 0.05 &   43.59 & \phn  2788   $\pm$\phn\phn  29 &\phn  1942   $\pm$\phn\phn  26 &   24 &  1\\ 
                                    &   43.06 $\pm$ 0.06 &   43.59 & \phn  3729   $\pm$\phn 426 &\phn  2168   $\pm$\phn\phn  30 &   17 &  1\\ 
                                    &   43.33 $\pm$ 0.05 &   43.59 & \phn  2743   $\pm$\phn\phn  79 &\phn  1967   $\pm$\phn\phn  19 &   16 &  1\\ 
                                    &   43.61 $\pm$ 0.08 &   43.59 & \phn  2500   $\pm$\phn\phn  43 &\phn  1880   $\pm$\phn\phn  19 &   17 &  1\\ 
3C120                               &   44.09 $\pm$ 0.09 &   43.19 & \phn  2327   $\pm$\phn\phn  48 &\phn  1249   $\pm$\phn\phn  21 &   52 &  1\\ 
Akn120                              &   43.95 $\pm$ 0.04 &   44.01 & \phn  6042   $\pm$\phn\phn  35 &\phn  1753   $\pm$\phn\phn\phn   6 &   20 &  1\\ 
                                    &   43.61 $\pm$ 0.06 &   44.01 & \phn  6246   $\pm$\phn\phn  78 &\phn  1862   $\pm$\phn\phn  13 &   20 &  1\\ 
Mrk79                               &   43.60 $\pm$ 0.06 &   43.80 & \phn  5056   $\pm$\phn\phn  85 &\phn  2314   $\pm$\phn\phn  23 &   20 &  1\\ 
                                    &   43.71 $\pm$ 0.06 &   43.80 & \phn  4760   $\pm$\phn\phn  31 &\phn  2281   $\pm$\phn\phn  26 &   19 &  1\\ 
                                    &   43.64 $\pm$ 0.06 &   43.80 & \phn  4766   $\pm$\phn\phn  71 &\phn  2312   $\pm$\phn\phn  21 &   23 &  1\\ 
                                    &   43.54 $\pm$ 0.05 &   43.80 & \phn  4137   $\pm$\phn\phn  37 &\phn  1939   $\pm$\phn\phn  16 &   24 &  1\\ 
PG0804+761                          &   44.88 $\pm$ 0.09 &   44.18 & \phn  3053   $\pm$\phn\phn  38 &\phn  1434   $\pm$\phn\phn  18 &   70 &  1\\ 
PG0844+349                          &   44.19 $\pm$ 0.06 &   43.73 & \phn  2694   $\pm$\phn\phn  58 &\phn  1505   $\pm$\phn\phn  14 &   48 &  1\\ 
Mrk110                              &   43.64 $\pm$ 0.06 &   42.65 & \phn  1543   $\pm$\phn\phn\phn   5 &\phn\phn   962   $\pm$\phn\phn  15 &   21 &  1\\ 
                                    &   43.72 $\pm$ 0.07 &   42.65 & \phn  1658   $\pm$\phn\phn\phn   3 &\phn\phn   953   $\pm$\phn\phn  10 &   14 &  1\\ 
                                    &   43.49 $\pm$ 0.15 &   42.65 & \phn  1600   $\pm$\phn\phn  39 &\phn\phn   987   $\pm$\phn\phn  18 &   28 &  1\\ 
PG0953+414                          &   45.15 $\pm$ 0.07 &   44.56 & \phn  3071   $\pm$\phn\phn  27 &\phn  1659   $\pm$\phn\phn  31 &   35 &  1\\ 
NGC3227                             &   42.86 $\pm$ 0.08 &   43.23 & \phn  4445   $\pm$\phn 134 &\phn  1914   $\pm$\phn\phn  71 &   23 &  1\\ 
                                    &   42.32 $\pm$ 0.05 &   43.23 & \phn  5103   $\pm$\phn 159 &\phn  2473   $\pm$\phn\phn  26 &   26 &  1\\ 
                                    &   42.11 $\pm$ 0.04 &   43.23 & \phn  3972   $\pm$\phn\phn  25 &\phn  1749   $\pm$\phn\phn\phn   4 &   75 &  2\\ 
NGC3516                             &   43.17 $\pm$ 0.15 &   43.77 & \phn  5236   $\pm$\phn\phn  12 &\phn  1584   $\pm$\phn\phn\phn   1 &   74 &  2\\ 
NGC3783                             &   43.02 $\pm$ 0.05 &   42.86 & \phn  3770   $\pm$\phn\phn  68 &\phn  1691   $\pm$\phn\phn  19 &   73 &  3\\ 
NGC4051                             &   41.88 $\pm$ 0.05 &   42.86 & \phn  1453   $\pm$\phn\phn\phn   3 &\phn  1500   $\pm$\phn\phn  34 &   29 & 1\\ 
                                    &   41.82 $\pm$ 0.03 &   42.86 & \phn\phn   799   $\pm$\phn\phn\phn   2 &\phn  1045   $\pm$\phn\phn\phn   4 &   86 & 4\\ 
PG1211+143                          &   44.70 $\pm$ 0.08 &   43.77 & \phn  2012   $\pm$\phn\phn  37 &\phn  1487   $\pm$\phn\phn  30 &   36 &  1\\ 
PG1226+023                          &   45.93 $\pm$ 0.07 &   45.05 & \phn  3509   $\pm$\phn\phn  36 &\phn  1778   $\pm$\phn\phn  17 &   39 &  1\\ 
PG1229+204                          &   43.65 $\pm$ 0.06 &   43.58 & \phn  3828   $\pm$\phn\phn  54 &\phn  1608   $\pm$\phn\phn  24 &   33 &  1\\ 
NGC4593                             &   42.85 $\pm$ 0.04 &   43.84 & \phn  5143   $\pm$\phn\phn  16 &\phn  1790   $\pm$\phn\phn\phn   3 &   25 & 5\\ 
PG1307+085                          &   44.82 $\pm$ 0.06 &   44.26 & \phn  5059   $\pm$\phn 133 &\phn  1963   $\pm$\phn\phn  47 &   23 &  1\\ 
IC4329A                             &   42.89 $\pm$ 0.07 &   44.36 & \phn  5964   $\pm$\phn 134 &\phn  2271   $\pm$\phn\phn  58 &   25 &  1\\ 
Mrk279                              &   43.66 $\pm$ 0.06 &   43.61 & \phn  5354   $\pm$\phn\phn  32 &\phn  1823   $\pm$\phn\phn  11 & $<40$ & 1\\ 
PG1411+442                          &   44.52 $\pm$ 0.05 &   44.08 & \phn  2801   $\pm$\phn\phn  43 &\phn  1774   $\pm$\phn\phn  29 &   24 &  1\\ 
NGC5548                             &   43.35 $\pm$ 0.07 &   43.93 & \phn  4674   $\pm$\phn\phn  63 &\phn  1934   $\pm$\phn\phn\phn   5 &  132 &  1\\ 
                                    &   43.09 $\pm$ 0.07 &   43.93 & \phn  5418   $\pm$\phn 107 &\phn  2227   $\pm$\phn\phn  20 &   94 & 1\\ 
                                    &   43.31 $\pm$ 0.06 &   43.93 & \phn  5236   $\pm$\phn\phn  87 &\phn  2205   $\pm$\phn\phn  16 &   65 & 1\\ 
                                    &   43.02 $\pm$ 0.09 &   43.93 & \phn  5986   $\pm$\phn\phn  95 &\phn  3110   $\pm$\phn\phn  53 &   83 & 1\\ 
                                    &   43.28 $\pm$ 0.06 &   43.93 & \phn  5930   $\pm$\phn\phn  42 &\phn  2486   $\pm$\phn\phn  13 &  142 & 1\\ 
                                    &   43.33 $\pm$ 0.06 &   43.93 & \phn  7378   $\pm$\phn\phn  39 &\phn  2877   $\pm$\phn\phn  17 &  128 & 1\\ 
                                    &   43.48 $\pm$ 0.05 &   43.93 & \phn  6946   $\pm$\phn\phn  79 &\phn  2432   $\pm$\phn\phn  13 &   78 & 1\\ 
                                    &   43.39 $\pm$ 0.08 &   43.93 & \phn  6623   $\pm$\phn\phn  93 &\phn  2276   $\pm$\phn\phn  15 &  144 & 1\\ 
                                    &   43.19 $\pm$ 0.06 &   43.93 & \phn  6298   $\pm$\phn\phn  65 &\phn  2178   $\pm$\phn\phn  12 &   95 & 1\\ 
                                    &   43.55 $\pm$ 0.06 &   43.93 & \phn  6177   $\pm$\phn\phn  36 &\phn  2035   $\pm$\phn\phn  11 &  119 & 1\\ 
                                    &   43.46 $\pm$ 0.08 &   43.93 & \phn  6247   $\pm$\phn\phn  57 &\phn  2021   $\pm$\phn\phn  18 &   86 & 1\\ 
                                    &   43.06 $\pm$ 0.08 &   43.93 & \phn  6240   $\pm$\phn\phn  77 &\phn  2010   $\pm$\phn\phn  30 &   37 & 1\\ 
                                    &   43.06 $\pm$ 0.07 &   43.93 & \phn  6478   $\pm$\phn 108 &\phn  3111   $\pm$\phn 131 &   45 & 1\\ 
                                    &   42.87 $\pm$ 0.05 &   43.93 & \phn  6396   $\pm$\phn 167 &\phn  3210   $\pm$\phn 642 &   28 & 6\\ 
PG1426+015                          &   44.60 $\pm$ 0.08 &   44.26 & \phn  7113   $\pm$\phn 160 &\phn  2906   $\pm$\phn\phn  80 &   20 &  1\\ 
Mrk817                              &   43.75 $\pm$ 0.07 &   42.78 & \phn  4711   $\pm$\phn\phn  49 &\phn  1984   $\pm$\phn\phn\phn   8 &   25 &  1\\ 
                                    &   43.63 $\pm$ 0.06 &   42.78 & \phn  5237   $\pm$\phn\phn  67 &\phn  2098   $\pm$\phn\phn  13 &   17 &  1\\ 
                                    &   43.63 $\pm$ 0.05 &   42.78 & \phn  4767   $\pm$\phn\phn  72 &\phn  2195   $\pm$\phn\phn  16 &   19 &  1\\ 
PG1613+658                          &   44.73 $\pm$ 0.07 &   44.62 & \phn  9074   $\pm$\phn 103 &\phn  3084   $\pm$\phn\phn  33 &   48 &  1\\ 
PG1617+175                          &   44.36 $\pm$ 0.09 &   44.11 & \phn  6641   $\pm$\phn 190 &\phn  2313   $\pm$\phn\phn  69 &   34 &  1\\ 
PG1700+518                          &   45.56 $\pm$ 0.05 &   44.83 & \phn  2252   $\pm$\phn\phn  85 &\phn  3160   $\pm$\phn\phn  93 &   37 &  1\\ 
3C390.3                             &   43.65 $\pm$ 0.08 &   43.62 &  12694   $\pm$\phn\phn  13 &\phn  3744   $\pm$\phn\phn  42 &  104 & 1\\ 
Mrk509                              &   44.16 $\pm$ 0.09 &   43.98 & \phn  3015   $\pm$\phn\phn\phn   2 &\phn  1555   $\pm$\phn\phn\phn   7 &  194 &  1\\ 
PG2130+099                          &   44.40 $\pm$ 0.05 &   42.94 & \phn  2853   $\pm$\phn\phn  39 &\phn  1485   $\pm$\phn\phn  15 &   21 & 7\\ 
NGC7469                             &   43.30 $\pm$ 0.04 &   44.04 & \phn  1722   $\pm$\phn\phn  30 &\phn  1707   $\pm$\phn\phn  20 &   54 & 1\\ 
\enddata

\tablecomments{Refs: (1)~\citet{peterson04} and references therein,
  (2)~\citet{denney10}, (3)~\citet{onken02}, (4)~\citet{denney09b},
  (5)~\citet{denney06}, (6)~\citet{bentz07}, (7)~\citet{grier08}. Mean
  line-widths for the spectra of \citet{peterson04} and
  \citet{onken02} are presented by \citet{collin06}.}

\end{deluxetable}

\end{document}